\documentclass[prc,aps,groupedaddress,superscriptaddress,twocolumn,nofootinbib,showpacs,showkeys,floatfix,a4paper,10pt,amsmath,amssymb]{revtex4-1}
\usepackage{graphicx}
\usepackage{dcolumn}
\usepackage{bm}
\usepackage[T1]{fontenc}

\begin{document}

\title{Impact of perturbative tensor interactions on the spontaneous fission half-lives of superheavy nuclei}

\author{R. Rodr\'{\i}guez-Guzm\'an}
\email{guzman.rodriguez@nu.edu.kz}
\affiliation{Department of Physics, School of Sciences and Humanities, Nazarbayev 
University, 53 Kabanbay Batyr Ave., Astana 010000, Kazakhstan}

\author{A. Rakhmankulov}
\email{aziz.rakhmankulov@nu.edu.kz}
\affiliation{Department of Physics, School of Sciences and Humanities, Nazarbayev 
University, 53 Kabanbay Batyr Ave., Astana 010000, Kazakhstan}

\author{L. M. Robledo}
\email{luis.robledo@uam.es}
\affiliation{%
Center for Computational Simulation, Universidad Polit\'ecnica de 
Madrid, Campus Montegancedo, 28660 Boadilla del Monte, Madrid, Spain
}%
\affiliation{Departamento  de F\'{\i}sica Te\'orica and CIAFF, 
Universidad Aut\'onoma de Madrid, 28049-Madrid, Spain}

\author{R. N. Bernard}
\email{remi.bernard@cea.fr}
\affiliation{%
CEA, DES, IRESNE, DER, SPRC, Cadarache, 13108, Saint-Paul-les-Durance, France
}%

\date{\today}

\begin{abstract} 
The standard microscopic description of fission, based on the mean-field 
Hartree-Fock-Bogoliubov approximation and a semi-classical description 
of tunneling through the fission barrier, has been used to analyse the 
impact of introducing a (perturbative) tensor term along with the well 
known Gogny-D1S force in the spontaneous fission half-lives. 
Calculations in a series of even-even  isotopes of superheavy nuclei 
ranging from nobelium to darmstatium  have been carried out. The 
results show that the tensor term only impacts the height of the first 
fission barrier and leaves mostly unaffected the pairing properties and 
therefore the collective inertias. As a consequence of the reduction in 
the barrier height, the spontaneous fission lifetimes obtained by 
including the tensor term are significantly smaller than the ones 
without it bringing the theoretical predictions in closer agreement 
with experimental data.
\end{abstract}


\maketitle{}

%
%
%

\section{Introduction}

Fission represents one of the main decay channels in heavy and 
superheavy nuclei 
\cite{turning,Specht-fission,Bjornholm-fission,Robledo-fission,SH-d1mstar}. 
Since its experimental discovery \cite{fission-dis} and initial 
interpretation \cite{Meitner-1,Bohr-1}, a lot of effort has been 
devoted to account for the most relevant dynamic correlations along the 
fission paths of atomic nuclei. Nuclear fission was originally 
described using the liquid-drop model, which emphasizes the competition 
between the nuclear surface tension and Coulomb repulsion 
\cite{Meitner-1,Bohr-1}. Nevertheless, quantum shell effects also play 
a significant role along the fission paths 
\cite{Ringbook,shelleffects-1,shelleffects-2}. Therefore, approaches 
incorporating those shells effects to the semi-classical liquid-drop 
have  been the subject of detailed theoretical studies 
\cite{liquid-shell-1,liquid-shell-2,liquid-shell-3,liquid-shell-4,liquid-shell-5}. 
Note that shell effects, key to the fission dynamics 
\cite{Robledo-fission,Kowal-1}, provide the only mechanism responsible 
for the stability of superheavy elements.

From a quantum mechanical perspective, spontaneous fission is usually 
considered as the transition of a given nuclear system from the ground 
state to the scission configuration precursor of the splitting into 
fission fragments. The transition mechanism requires the introduction 
of an energy landscape parametrized using  intrinsic shapes labeled by 
several deformation parameters $\bf{Q}$ that characterize the shape 
evolution of the nucleus as it evolves from the initial configuration 
to the fissioned one. In the last decades, the microscopic 
$\bf{Q}$-constrained mean-field approximation based on the 
Hartree-Fock-Bogoliubov
(HFB) method 
has emerged  as a valuable tool, complementary to other theoretical 
approaches, to describe the fission landscape \cite{Robledo-fission}. 
The $\bf{Q}$-constrained  mean-field approximation provides the optimal 
fission configurations (minima, valleys, ridges and saddle points), as 
well as the associated pairing  correlations and shell effects. It uses 
the  Ritz variational minimization principle \cite{Ringbook} of an 
energy density functional (EDF) in the space of HFB wave functions with 
constrains on shape multipole operators such as the axial 
$\hat{Q}_{20}$ and triaxial $\hat{Q}_{22}$ quadrupole, octupole 
$\hat{Q}_{30}$, necking $\hat{Q}_{Neck}$, $\cdots$ operators, i.e., 
${\bf{Q}}=\Big(Q_{20},Q_{22},Q_{30},Q_{Neck}, \cdots \Big)$. It also 
provides quantities such as collective inertias $M(\bf{Q})$ and 
zero-point rotational $E_{ROT}(\bf{Q})$ and vibrational 
$E_{VIB}(\bf{Q})$ energy corrections. These building blocks allow the 
computation of the spontaneous fission half-lives $t_{SF}$ using the 
Wentzel-Kramers-Brillouin (WKB) formalism \cite{Rayner-U-PRC,WKB-1, 
WKB-2}.

\begin{figure*}
\includegraphics[width=1.0\textwidth]{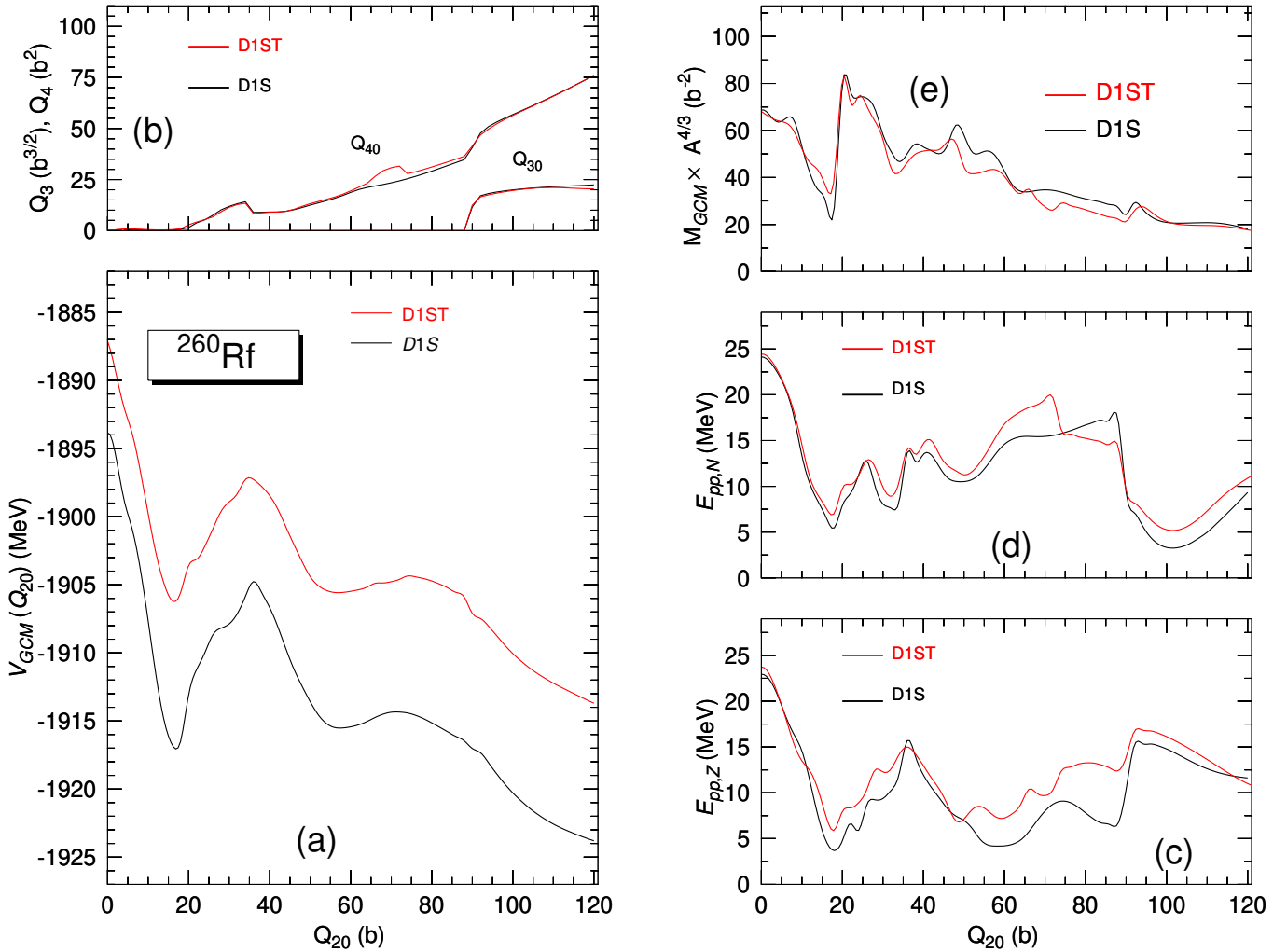}
\caption{(Color online) The GCM collective  potentials 
$V_{GCM}(Q_{20})$ Eq.(\ref{Coll-Pot-V-Q20}) obtained for the nucleus 
$^{260}$Rf are plotted in panel (a) as functions of the quadrupole 
moment $Q_{20}$. The octupole $Q_{30}$ and hexadecapole $Q_{40}$ 
moments of the intrinsic states are plotted in panel (b). The proton 
$E_{pp,Z}$ and neutron $E_{pp,N}$ pairing interaction energies are 
depicted in panels (c) and (d), while the GCM collective  mass is 
plotted in panel (e). Results have been obtained with the 
parametrizations D1S and D1ST of the Gogny-EDF. For more details, see 
the main text. 
}
\label{Pegagogival_260Rf} 
\end{figure*} 

Microscopic $\bf{Q}$-constrained mean-field  fission calculations
have already provided a wealth of information on fission properties
in heavy and superheavy nuclei
\cite{SH-d1mstar,Rayner-U-PRC,MF-fission-1,MF-fission-2,MF-fission-3,MF-fission-4,MF-fission-5,MF-fission-6,MF-fission-7,MF-fission-8,MF-fission-9,MF-fission-10,MF-fission-11,MF-fission-12,MF-fission-13,MF-fission-14,MF-fission-15,MF-fission-16,MF-fission-17,MF-fission-18,MF-fission-19,MF-fission-20,MF-fission-21,MF-fission-22,MF-fission-23}.
In addition to the traditional fission disintegration mode, cluster radioactivity has also
been  described as a super-asymmetric fission mode employing
the $\bf{Q}$-constrained  mean-field framework 
\cite{cluster-ra-1,cluster-ra-2,cluster-ra-3,cluster-ra-4,cluster-ra-5}. 

The role of pairing correlations in the fission dynamics has also 
received  attention within the least action framework where the path to 
fission followed in the space of collective coordinates is determined 
by minimizing the action integral associated to the collective degrees 
of freedom. Both non-relativistic and relativistic studies 
\cite{MA-1,MA-2,MA-3,MA-4,MA-5,MA-6,MA-7,MA-8,MA-9,MA-10} have shown, 
that the action is severely quenched when pairing degrees of freedom 
are included as collective degrees of freedom in the minimization 
procedure. As a consequence, the  predicted $t_{SF}$ values get a 
reduction of several orders of magnitude, improving  substantially the 
comparison with experimental $t_{SF}$ values \cite{exp-tsf}. For an 
account of the role of dynamic pairing correlations using a 
(restricted) particle number variation-after-projection (VAP) approach 
to fission, the reader is referred to 
Refs.~\cite{RVAP-PNP-1,RVAP-PNP-2}.

\begin{figure*}
\includegraphics[width=0.98\textwidth]{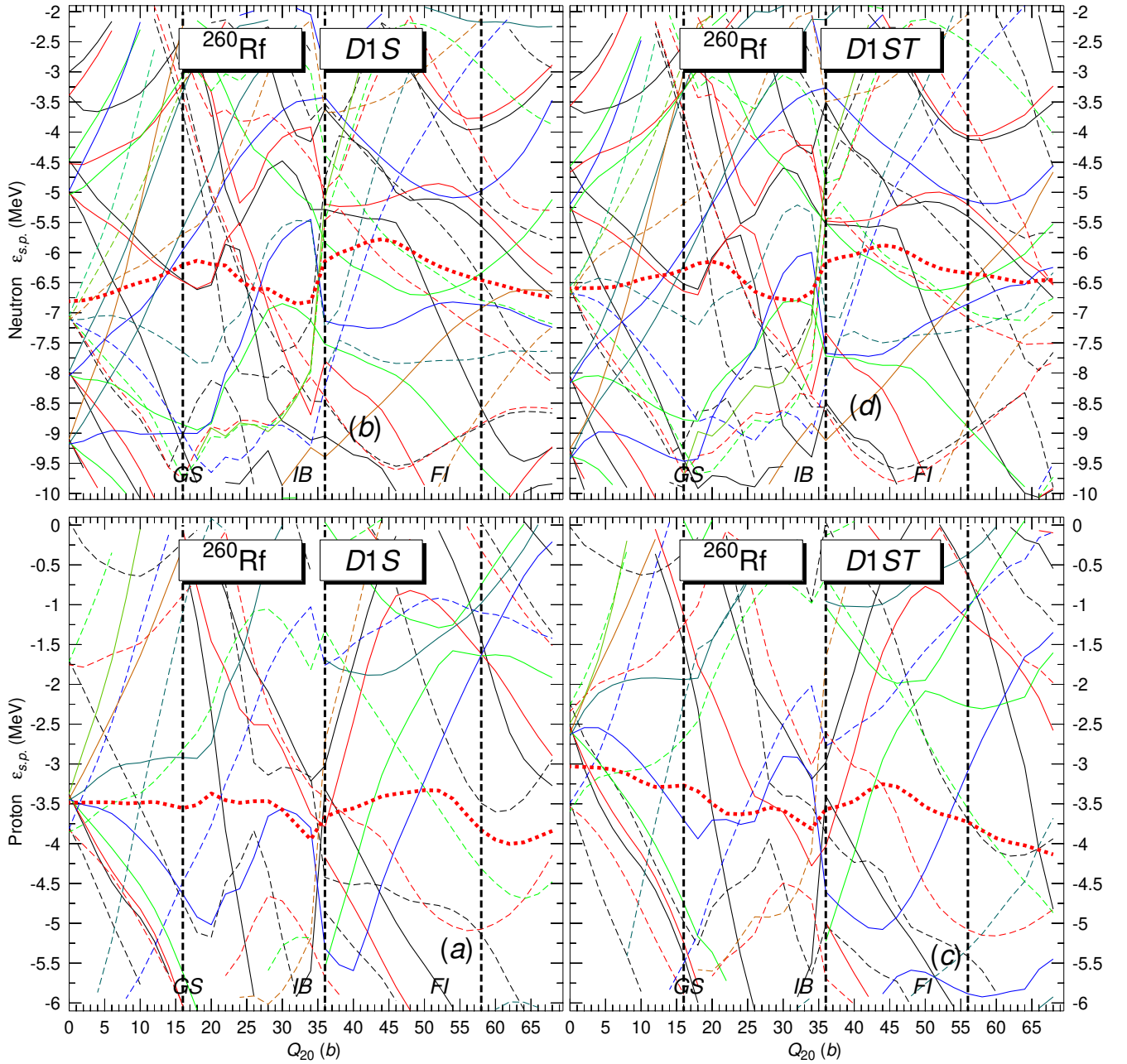}
\caption{(Color online) The proton [panel (a)] and neutron [panel (b)] 
single-particle energies obtained for the nucleus $^{260}$Rf with the 
parametrization D1S of the Gogny-EDF are depicted as functions of the 
quadrupole moment $Q_{20}$. The D1ST proton and neutron single-particle 
energies are shown in panels (c) and (d). The Fermi levels are plotted 
with a thick (red) dotted line. Solid (dashed) lines are used for 
positive (negative) parity states. With increasing $K =1/2, 3/2, 5/2, \ldots $ 
values color labels are black, red, green, blue, dark-blue, brown, 
dark-green, etc. To guide the eye, vertical thick (black) dashed lines 
are shown at the quadrupole deformations corresponding to the ground 
state (GS), the top of the inner barrier (IB) and the fission isomer 
(FI). For more details, see the main text.
}
\label{spes_260Rf} 
\end{figure*}

Despite significant advances in the theoretical computational setup, 
achieving an accurate description of the fission energy landscape 
continues to be a major challenge in  nuclear structure physics 
nowadays. There is still a vivid discussion not only on the many-body 
methods  but also on the type and performance of phenomenological 
interactions applied to describe fission \cite{Robledo-fission}.

Mean-field fission calculations usually neglect the tensor component of 
the nuclear interaction associated to the one-pion exchange long range 
potential and its impact on the single particle distribution around the 
Fermi level (i.e. its impact on shell effects). However,  tensor 
correlations in nuclear structure have been the subject of renewed 
interest within the non-relativistic and relativistic EDF frameworks 
\cite{Tensor-Stancu,Tensor-pert-Skyrme-1,Tensor-pert-Skyrme-2,Tensor-pert-Skyrme-3, 
Tensor-pert-Skyrme-4,Tensor-pert-Skyrme-5,Tensor-pert-Skyrme-6, 
Tensor-Skyrme-1,Tensor-Skyrme-2,Tensor_Gogny-Otsuka,Otsuka25,Marta-tensor-1,Marta-tensor-2,Marta-tensor-3,Marta-tensor-4, 
Marta-tensor-5,finite-range-tensor-refit-gogny,MY3-1,MY3-2,MY3-3,tensor-RMF-1,tensor-RMF-2,tensor-RMF-3,tensor-RMF-4}. 
In those previous studies, tensor effects have been  accounted for  
perturbatively  (on top of existing parametrizations) or via a full 
refit of the EDF parameters.

The Gogny-D1 family of effective interactions \cite{Review-Rayner} 
includes parametrizations such as D1S \cite{MF-fission-1}, D1N 
\cite{gogny-d1n}, D1M \cite{gogny-d1m} and D1M$^{*}$ 
\cite{gogny-d1mstar}. The D1S parametrization is considered a  standard 
within the family, while D1N, D1M and D1M$^{*}$ have brought a more 
accurate description of nuclear binding energies  and neutron star  
masses. These parametrizations have already shown a reasonable 
predictive power when applied to describe fission properties in heavy 
and superheavy nuclei 
\cite{SH-d1mstar,Rayner-U-PRC,MF-fission-1,MF-fission-2,MF-fission-3,MF-fission-4,MF-fission-5,MF-fission-6,MF-fission-7,MA-1,MA-3,MA-4,RVAP-PNP-1,RVAP-PNP-2,Review-Rayner}. 
The parametrization D1ST2a (to be referred simply as D1ST hereafter) 
has been obtained via the perturbative addition of a complete long 
range tensor term to the well tested Gogny-D1S EDF 
\cite{Marta-tensor-1,Marta-tensor-2,Marta-tensor-3,Marta-tensor-4,Marta-tensor-5}.

Two previous studies  
\cite{Fission-tensor-2020,Fission-tensor-raynerJPG}, based on the 
Gogny-D1ST EDF, have provided  valuable initial insights into the role 
of tensor correlations in microscopic Gogny-HFB
fission calculations. In the case of neutron deficient Th 
isotopes, it has been found  \cite{Fission-tensor-2020} that the impact 
of the tensor interaction is crucial to account for  a new fission 
valley associated with exotic symmetric bimodal fission  
\cite{exp-nd-Th,bimodal-1,bimodal-2}. An account of the consequences of 
adding perturbatively the tensor interaction  on the spontaneous 
fission half-lives of the heavy actinides $^{240-250}$Cm and 
$^{240-250}$Cf as well as  the low-Z superheavy nuclei $^{242-258}$Fm 
was presented in Ref.~\cite{Fission-tensor-raynerJPG}. It has been 
found, that the Gogny-D1ST inner fission barrier heights are 
significantly reduced as compared to the ones obtained in Gogny-D1S 
calculations (i.e., without the tensor). For the considered Cm, Cf and 
Fm nuclei, the Gogny-D1ST reduction of the inner barrier heights 
represents a key tensor-driven mechanism with a significant impact on 
the predicted $t_{SF}$ values \cite{Fission-tensor-raynerJPG}.

\begin{figure}
\includegraphics[width=0.45\textwidth]{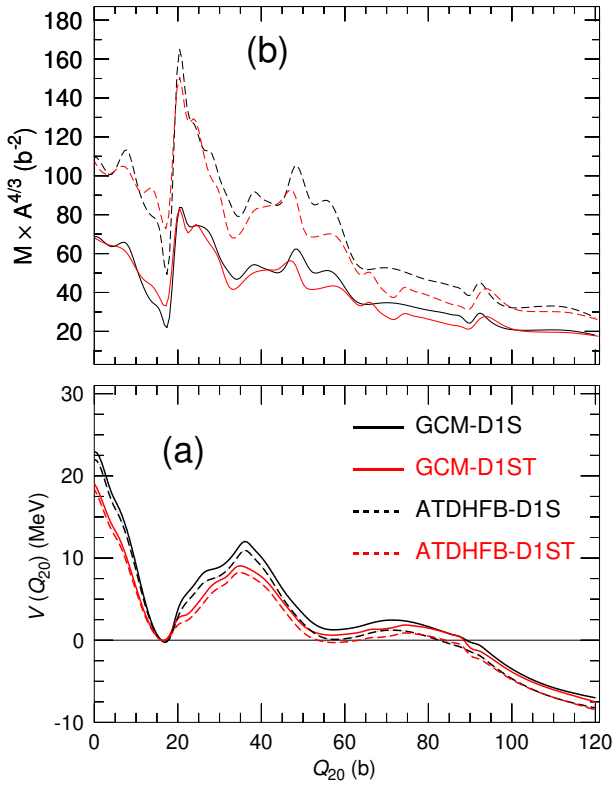}
\caption{(Color online) The collective  potentials $V(Q_{20})$ 
Eq.(\ref{Coll-Pot-V-Q20}),  obtained within the GCM and ATDHFB schemes, 
for the nucleus $^{260}$Rf, are plotted in panel (a) as functions of 
the quadrupole moment $Q_{20}$. All the relative energies are measured 
with respect to the absolute minima of the corresponding paths. The GCM 
and ATDHFB collective masses are plotted in panel (b). Results have 
been obtained with the parametrizations D1S and D1ST of the Gogny-EDF. 
For more details, see the main text.
}
\label{comp-GCM-ATD} 
\end{figure}

Certainly, much work is still needed to deepen our understanding of how 
correlations induced by the perturbative tensor interaction influence 
fission-related properties. In this respect, it has to be mentioned 
that the main reason driving the choice of the nuclei $^{240-250}$Cm, 
$^{240-250}$Cf and $^{242-258}$Fm in our previous study 
\cite{Fission-tensor-raynerJPG} was that they belong to a region of the 
Segr\'e chart  where the transition from the actinides to superheavy 
nuclei takes place and the second barrier height decreases 
substantially \cite{SH-d1mstar}. On the other hand, superheavy elements 
have been the subject of intense experimental effort 
\cite{SH-EXP-effort-1,SH-EXP-effort-2,SH-EXP-effort-3} as their 
stability is closely related to spontaneous fission properties 
\cite{Cwiok-1,Nazarew-1}. They are also interesting because the second
barrier is nearly gone in many of them and therefore fission is only driven 
by the physics at the first barrier. Therefore, it is timely and necessary to 
extend our previous study \cite{Fission-tensor-raynerJPG} to the region 
of superheavy nuclei with atomic number $Z > 100$ to disentangle the 
impact of tensor contributions on their spontaneous fission lifetimes. 
To the best of our knowledge, at least in the case of Gogny-like 
interactions, such an account is not yet available in the literature. 
This is precisely the aim of this work where, attention is paid to the 
spontaneous fission properties of a selected set of even-even No, Rf, 
Sg, Hs and Ds nuclei via the comparison between Gogny-D1S and 
Gogny-D1ST results obtained using the HFB framework 
\cite{Ringbook,Review-Rayner}.

The paper is organized as follows. In Sec.~\ref{Theory}, we briefly 
outline the constrained HFB method employed in the calculation of the 
mean field configurations characterizing the potential energy surfaces 
and collective inertias \cite{Review-Rayner,Fission-tensor-raynerJPG}. 
The results of such calculations for $^{250-266}$No, $^{254-270}$Rf, 
$^{256-278}$Sg,$^{258-280}$Hs and $^{264-282}$Ds are discussed in 
Sec.~\ref{results}. The discussion starts with 
Sec.~\ref{methodology-260Rf}, where the methodology employed to compute 
the Gogny-D1S and Gogny-D1ST fission paths and other fission-related 
quantities in the case of $^{260}$Rf is illustrated. The same 
methodology has been employed for all the nuclei studied in this paper. 
The systematic of the fission paths and spontaneous fission half-lives 
$t_{SF}$ is discussed in the next section Sec.~\ref{systematic-FP}. 
Besides the comparison with the available experimental $t_{SF}$ values 
\cite{exp-tsf}, for the studied isotopic chains this section examines 
the robustness of the predicted $t_{SF}$ trends, as functions of the 
neutron number, with respect to the different theoretical schemes 
employed to compute the collective potential and inertias. Finally, 
Sec.~\ref{conclusions} is devoted to the concluding remarks and work 
perspectives.

%

\section{Theoretical framework}
\label{Theory}
The results of constrained 
Gogny-D1S \cite{MF-fission-1}  and Gogny-D1ST
\cite{Marta-tensor-1,Marta-tensor-2,Marta-tensor-3,Marta-tensor-4,Marta-tensor-5}
HFB calculations are compared in order to disentangle 
the role of the tensor interaction in the spontaneous fission properties of 
a selected set 
of even-even superheavy No, Rf, Sg, Hs and Ds nuclei. The Gogny-D1S EDF is already well
known in the literature \cite{MF-fission-1,Review-Rayner}. In the case 
of the Gogny-D1ST EDF, the parameters of the 
finite-range central as well as the  zero-range two-body spin-orbit 
and density dependent terms are the same as in the D1S case
\cite{MF-fission-1,Marta-tensor-1,Marta-tensor-2,Marta-tensor-3,Marta-tensor-4,Marta-tensor-5}. 
What marks the difference is the introduction of a finite, long range ($\mu_{TS} =$ 1.2 fm) tensor term 
\begin{equation}
\label{tensor_D1ST}
V_{\textbf Tensor}(\vec{r}) =
\Big(V_{T1}+ V_{T2} P_{12}^{\tau}\Big) S_{12}(\vec{r})
e^{-\frac{\vec{r}^{2}}{\mu_{TS}^{2}}}
\end{equation}
added perturbatively. In the above expression, $\vec{r}$
represents the relative coordinate, $P_{12}^{\tau}$ stands for 
the isospin exchange operator, while the form factor $S_{12}(\vec{r})$
reads
\begin{equation}
S_{12}(\vec{r}) = 3 ~
\frac{\Big(\vec{\sigma}_{1} \cdot \vec{r} \Big) \Big(\vec{\sigma}_{2} \cdot \vec{r} \Big)}{r^{2}}
- \vec{\sigma}_{1} \cdot \vec{\sigma}_{2}
\end{equation}
with $\vec{\sigma}_{i}$ $(i=1,2)$ being the set of Pauli spin matrices. The 
parameters $V_{T1}$ and $V_{T2}$ have been  
adjusted to reproduce the $1 f_{5/2}$ and $1 f_{ 7/2}$  neutron 
single-particle energies in $^{48}$~Ca \cite{Marta-tensor-1,Marta-tensor-2,Marta-tensor-3,Marta-tensor-4,Marta-tensor-5}.

\begin{figure*}
\includegraphics[width=1.0\textwidth]{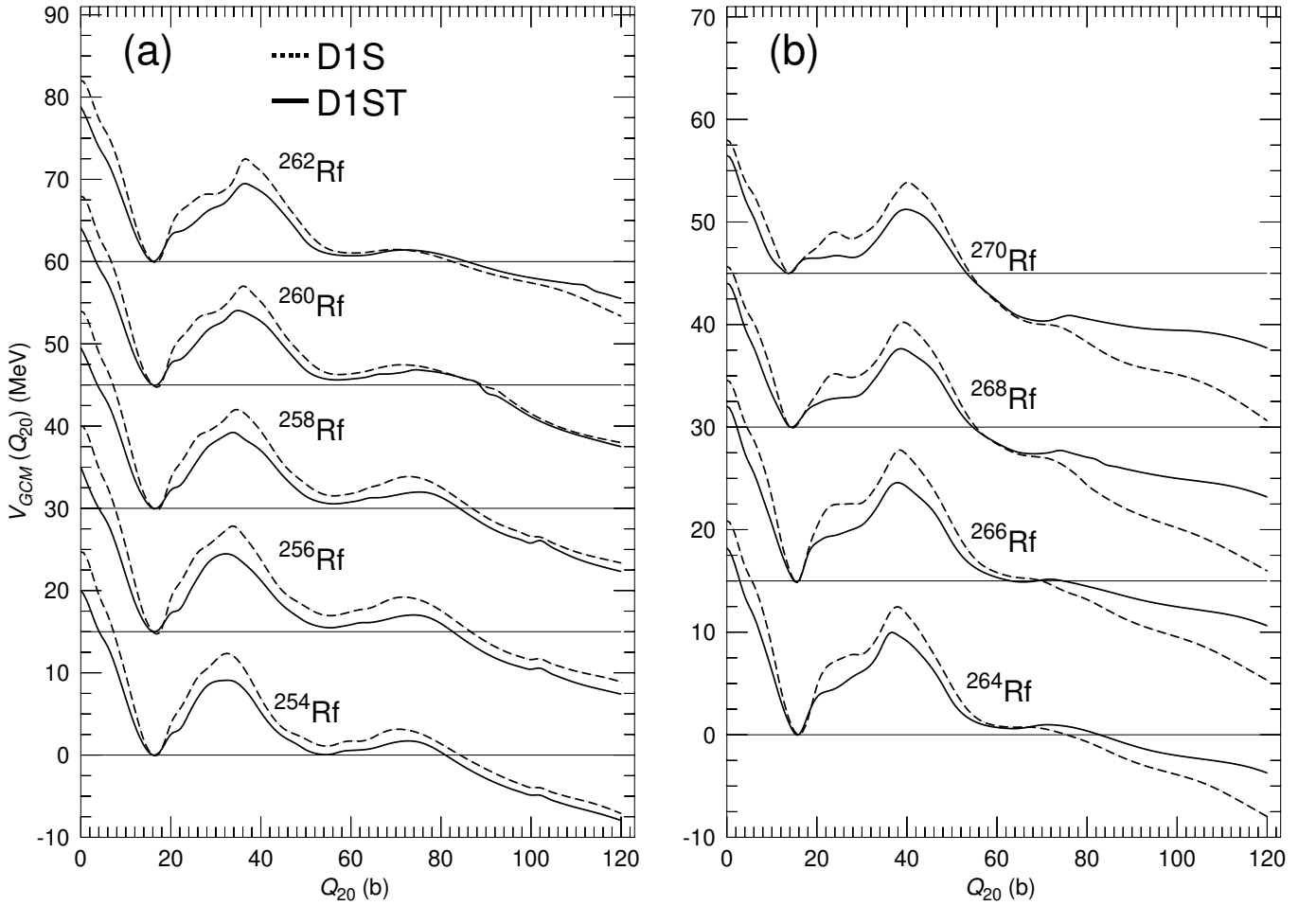}
\caption{The GCM collective  potentials $V_{GCM}(Q_{20})$ Eq.(\ref{Coll-Pot-V-Q20}) obtained 
for the nuclei $^{254-270}$Rf, are plotted as functions of the 
quadrupole moment $Q_{20}$. All the relative energies are measured with 
respect to the absolute minima of the corresponding paths. 
Starting from $^{256}$Rf ($^{266}$Rf) in
panel (a) [in panel (b)] all the curves have been successively shifted by 15 MeV in order to 
accommodate them in a single plot. Results have 
been obtained with the parametrizations D1S and D1ST of the Gogny-EDF. 
For more details, see the main text.
}
\label{FissionPAths_Rf} 
\end{figure*}

In all the constrained HFB calculations, Coulomb exchange is treated in 
the Slater approximation \cite{CSlater} while Coulomb and spin-orbit 
anti-pairing are neglected. The two-body kinetic energy correction has 
been fully taken into account in the selfconsistent procedure 
\cite{gradient-second} used to solve  the  constrained HFB equations. 
For the Gogny-D1ST EDF the contribution of the tensor term 
Eq.(\ref{tensor_D1ST}) to the pairing field has been neglected 
\cite{Marta-tensor-1,Marta-tensor-2,Marta-tensor-3,Marta-tensor-4,Marta-tensor-5, 
Fission-tensor-2020,Fission-tensor-raynerJPG}. Both axial and simplex 
symmetries have been kept as selfconsistent symmetries \cite{Ringbook} 
to alleviate the numerical effort.

We have employed a deformed axially symmetric harmonic oscillator (HO) 
basis containing states with $J_{z}$ quantum numbers up to 35/2 and up 
to 26 quanta in the z-direction. The basis quantum numbers 
$(n_{z},n_{\perp},m)$ are restricted by the condition 
\begin{equation} 2
n_{\perp} +|m| + \frac{1}{q} n_{z} \le N_{0} 
\end{equation} 
with $N_{0}=17$ and $q=1.5$. We have also optimized 
the two HO lengths b$_{\perp}$ and b$_{z}$ for each configuration along 
the fission paths of the studied nuclei 
\cite{SH-d1mstar,Rayner-U-PRC,Fission-tensor-raynerJPG}.  

Both Gogny-D1S and Gogny-D1ST  HFB calculations have been  carried out 
with an explicit (driving) constrain on the quadrupole 
\cite{Fission-tensor-raynerJPG}
\begin{equation}
\hat{Q}_{20} = z^{2} - \frac{1}{2}\Big (x^{2} + y^{2})
\end{equation}
operator. The quadrupole moment  of each intrinsic HFB state $| \varphi 
\rangle$ is  computed as the average value $Q_{20} = \langle \varphi | 
\hat{Q}_{20} | \varphi \rangle$. Let us stress,  that spatial 
reflection symmetry can be broken in the
calculations, whenever octupole shapes are favored energetically, to 
allow for the appearance of mass asymmetric 
fission paths.
  Due to this, we 
have also resorted to a constrain on the $\hat{Q}_{10}$  operator to  
prevent spurious effects associated to the center of mass motion 
\cite{2DGCM-q2q3-Gogny-3,2DGCM-q2q3-Gogny-4,2DGCM-q2q3-Gogny-5}.

\begin{figure}
\includegraphics[width=0.42\textwidth]{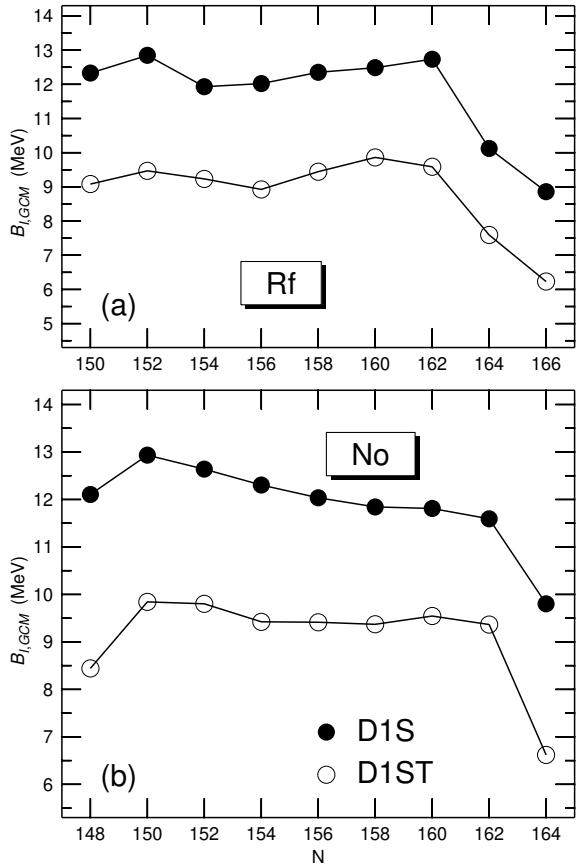}
\caption{Inner barrier heights $B_{I,GCM}$ for the nuclei $^{254-270}$Rf [panel (a)]
and $^{250-266}$No [panel (b)], as functions of the neutron number. Note, that the 
$B_{I,GCM}$ values contain the effects associated with zero-point  
rotational and vibrational corrections. Results have 
been obtained with the parametrizations D1S and D1ST of the Gogny-EDF.
For more details, see the main text.
}
\label{Inner_Rf_No} 
\end{figure}

As a result of the constrained HFB calculations, we have obtained a set 
of (intrinsic)  energies $E_{HFB}(Q_{20})$ and states $| \varphi 
(Q_{20}) \rangle$ in a  $Q_{20}$-mesh with $Q_{20} \in [0,120] ~b$ and 
the step size $\delta Q_{20}= 1 b$. We have checked that this 
$Q_{20}$-mesh is accurate enough to describe the fission properties of 
the studied nuclei. 

The zero-point rotational $E_{ROT}(Q_{20})$ and vibrational 
$E_{VIB}(Q_{20})$ energy corrections have been subtracted {\it{a 
posteriori}} from the HFB energies to obtain the collective potentials 
\begin{equation} 
\label{Coll-Pot-V-Q20}
V(Q_{20}) = E_{HFB}(Q_{20})- E_{VIB}(Q_{20}) 
- E_{ROT}(Q_{20}) 
\end{equation}
The rotational $E_{ROT}(Q_{20})$ correction has been computed in terms 
of the Yoccoz moment of inertia \cite{exact-rota-Rayner}. We have used 
a prescription \cite{RET-Reinhard} that correctly accounts for the fact 
that the $Q_{20}=0$  configuration is already a pure $0^{+}$ state, 
i.e., the rotational correction must satisfy $E_{ROT}(Q_{20}=0)=0$ 
\cite{exact-rota-Rayner,RET-Reinhard}.

Most of the results explicitly discussed in this paper, have been 
obtained with collective masses $M(Q_{20})$ and vibrational 
$E_{VIB}(Q_{20})$ corrections  computed within the (perturbative 
cranking) Gaussian Overlap  Approximation (GOA) to the Generator 
Coordinate Method (GCM). However, in some instances we will also 
compare with results obtained using collective masses and vibrational 
corrections computed within the (perturbative cranking) Adiabatic Time 
Dependent HFB (ATDHFB) scheme 
\cite{Rayner-U-PRC,collective-massesSamuel}. Therefore, for each 
nucleus we have computed the GCM collective potential $V_{GCM}(Q_{20})$ 
and mass $M_{GCM}(Q_{20})$ as well as the ATDHFB ones 
$V_{ATDHFB}(Q_{20})$ and $M_{ATDHFB}(Q_{20})$, as functions of the 
quadrupole deformation.

\begin{figure*}
\includegraphics[width=1.0\textwidth]{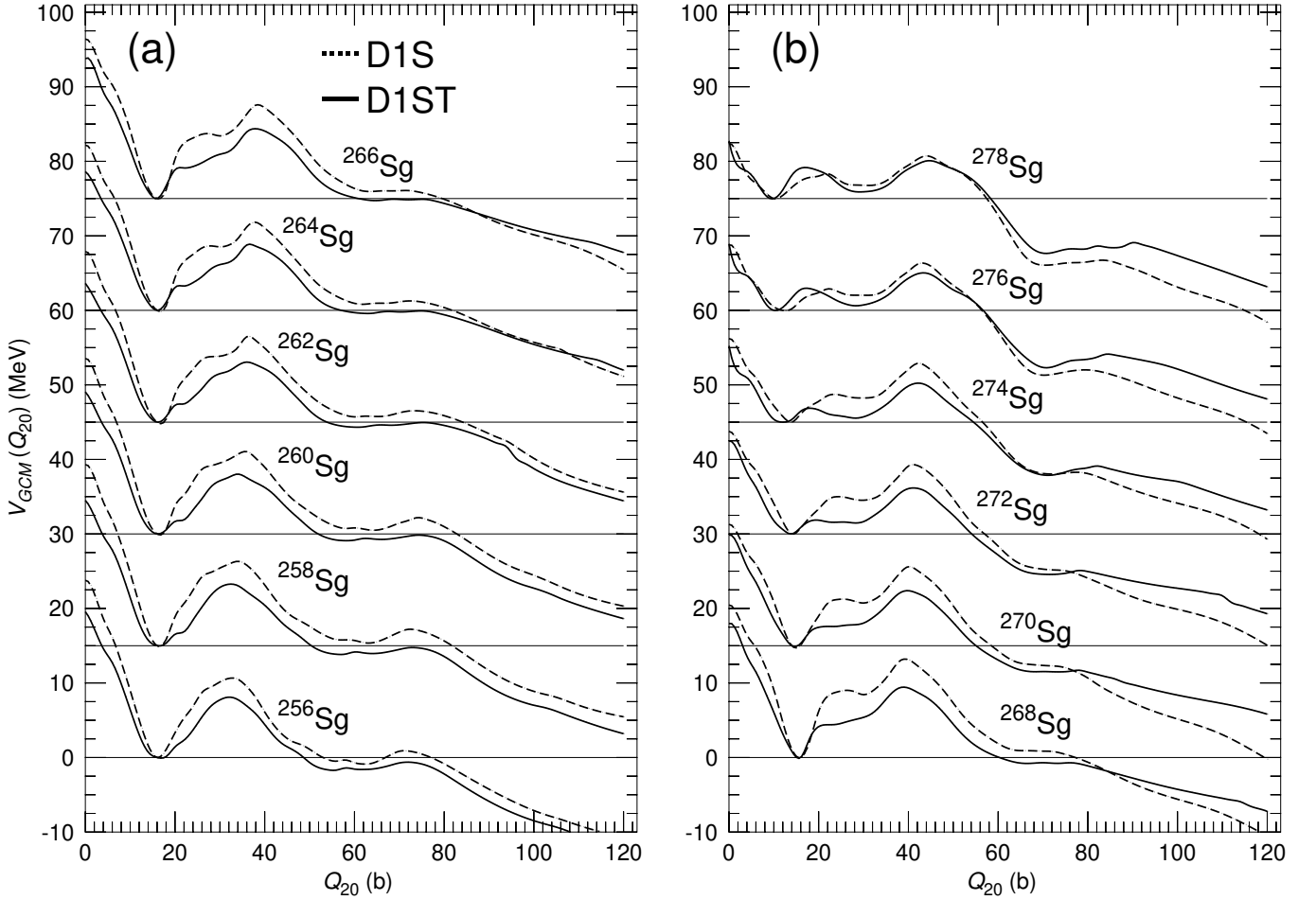}
\caption{The GCM collective  potentials $V_{GCM}(Q_{20})$ Eq.(\ref{Coll-Pot-V-Q20}) obtained 
for the nuclei $^{256-278}$Sg, are plotted as functions of the 
quadrupole moment $Q_{20}$. All the relative energies are measured with 
respect to the absolute minima of the corresponding paths. 
Starting from $^{258}$Sg ($^{270}$Sg) in
panel (a) [in panel (b)] all the curves have been successively shifted by 15 MeV in order to 
accommodate them in a single plot. Results have 
been obtained with the parametrizations D1S and D1ST of the Gogny-EDF. 
For more details, see the main text.
}
\label{FissionPAths_Sg} 
\end{figure*}

We have obtained
the spontaneous fission half-lives $t_\textrm{SF}$ (in 
seconds) using the WKB formalism \cite{Rayner-U-PRC,WKB-1, WKB-2}
\begin{equation} \label{TSF}
t_\mathrm{SF}= 2.86 \times 10^{-21} \times \left(1+ e^{2S} \right)
\end{equation}
where the action $S$ reads
\begin{equation} \label{Action}
S= \int_{a}^{b} dQ_{20} S(Q_{20})
\end{equation}
with the integrand 
\begin{equation}
\label{integrand-eq}
S(Q_{20}) = \sqrt{2M(Q_{20})\left(V(Q_{20})-\left(E_{Min}+E_{0} \right)  \right)}
\end{equation}
The integration limits $a$ and $b$ in Eq.(\ref{Action}) correspond to 
classical turning points \cite{turning} evaluated for the energy 
$E_{Min} + E_{0}$. The energy $E_{Min}$ corresponds to the absolute 
minimum of the considered path while, as in our previous study 
\cite{Fission-tensor-raynerJPG}, we have adopted the typical value 
$E_{0} = 0.5$ MeV for the true ground state energy once quadrupole 
fluctuations are taken into account.

%


\section{Discussion of the results} 
\label{results} 
In this section, we discuss the results of the calculations for 
$^{250-266}$No, $^{254-270}$Rf, $^{256-278}$Sg,$^{258-280}$Hs and 
$^{264-282}$Ds. First, in Sec.~\ref{methodology-260Rf}, we discuss in 
detail the followed methodology and the obtained results in the case of 
$^{260}$Rf, taken as a representative example. In 
Sec.~\ref{systematic-FP}, we discuss the systematic of the D1S and D1ST 
fission paths and spontaneous fission half-lives in the considered No, 
Rf, Sg, Hs and Ds nuclei.

\begin{figure*}
\includegraphics[width=1.0\textwidth]{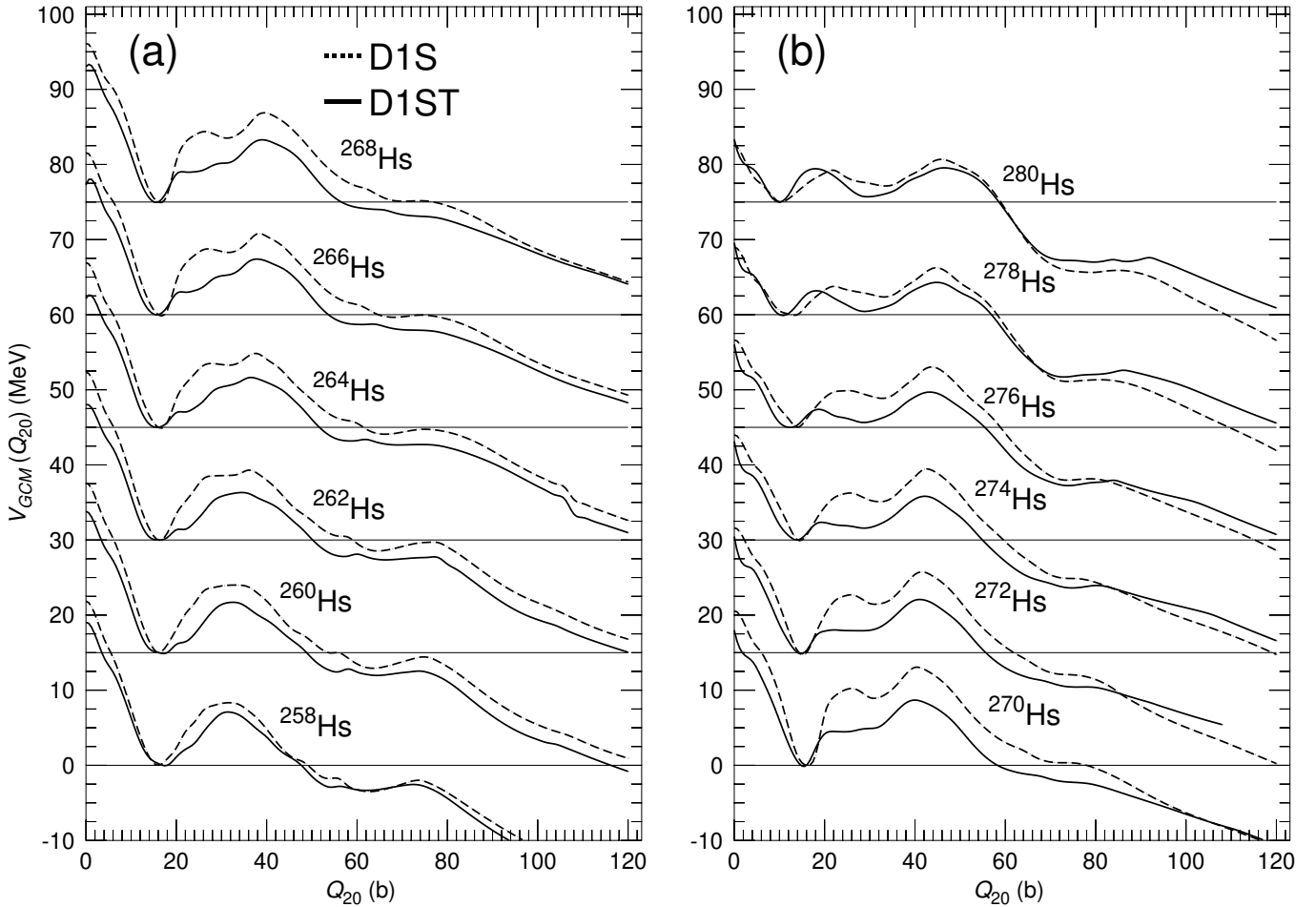}
\caption{The GCM collective  potentials $V_{GCM}(Q_{20})$ Eq.(\ref{Coll-Pot-V-Q20}) obtained 
for the nuclei $^{258-280}$Hs, are plotted as functions of the 
quadrupole moment $Q_{20}$. All the relative energies are measured with 
respect to the absolute minima of the corresponding paths. 
Starting from $^{260}$Hs ($^{272}$Hs) in
panel (a) [in panel (b)] all the curves have been successively shifted by 15 MeV in order to 
accommodate them in a single plot. Results have 
been obtained with the parametrizations D1S and D1ST of the Gogny-EDF. 
For more details, see the main text.
}
\label{FissionPAths_Hs} 
\end{figure*}


\subsection{An illustrative example: the nucleus $^{260}$Rf}
\label{methodology-260Rf}

The GCM collective potentials $V_{GCM}(Q_{20})$ 
Eq.(\ref{Coll-Pot-V-Q20}) obtained for $^{260}$Rf are plotted in panel 
(a) of Fig.~\ref{Pegagogival_260Rf}, as functions of the quadrupole 
moment $Q_{20}$ of the intrinsic states. The D1ST and D1S potentials 
exhibit a similar behavior with features such as ground state minima, 
inner barriers, fission isomers and outer barriers. From panel (b), one  
realizes that the octupole $Q_{30}$ and hexadecapole $Q_{40}$ 
deformations of the intrinsic states are also  similar along the 
fission paths obtained with both parametrizations of the Gogny-EDF. 
Nevertheless, the D1ST potential in panel (a) displays a pronounced 
under-binding relative to the D1S one. Such an under-binding is also 
found for all the nuclei considered in this work. As already pointed 
out in previous 
studies~\cite{Fission-tensor-2020,Fission-tensor-raynerJPG}, this 
effect originates from the perturbative character of D1ST, i.e., the tensor 
term is not taken into account in the fitting protocol that includes, 
among others, binding energies of some benchmark nuclei. 

For both GCM potentials, the  ground state and the top of the inner 
barrier are located at $Q_{20}=16$ and  $36$~b, respectively. The D1S 
(D1ST) fission isomer and the top of the outer barrier are located at 
$Q_{20}=58$~b ($56$~b) and $72$~b ($74$~b). For $^{260}$Rf, the fission 
isomer is reflection symmetric as octupole correlations start to play a 
role for quadrupole deformations around $Q_{20}=86$~b [see, panel (b)].

Notwithstanding their similar  behavior, the collective potentials 
differ significantly in the height $B_{I,GCM}$ of their inner barriers. 
We have obtained the  values $B_{I,GCM}^{D1S}=12.02$ and 
$B_{I,GCM}^{D1ST}=8.93$~MeV. In our calculations, axial symmetry has 
been preserved as a selfconsistent symmetry throughout the entire 
fission paths of the studied nuclei. The effect of the $\gamma$ 
deformation cannot be included at the present stage along with the 
tensor force, due to the lack of the corresponding computational code. 
  
It is well known that triaxiality can  decrease the inner barrier 
heights by a few MeV \cite{Rayner-U-PRC, MF-fission-2, MF-fission-16}. 
However, this reduction in the inner  barrier height comes together 
with an increase in the collective mass, which partially compensates 
for the change in the action, Eqs.(\ref{Action}) and 
(\ref{integrand-eq}). As a result, triaxiality plays a minor role in 
the predicted $t_{SF}$ values \cite{MF-fission-12, MF-fission-21}. 
Furthermore, it is worth noting that previous least action studies have 
shown that incorporating dynamical pairing fluctuations can fully 
restore axial symmetry along the fission path \cite{MA-7, MA-8}.
On the other hand, it has recently been argued \cite{Otsuka25} that the
tensor interaction might favour the development of triaxiality in some
rare earth nuclei.  Plotted as a function of the $\gamma$ deformation
parameter the tensor contribution to the energy shows in some cases a
minimum at around $\gamma =15$ that is compensated by other terms in the
central part of the interaction. One may wonder what would be the impact
of the tensor term in the expected reduction of the barrier height due
to triaxiality. This is not a trivial question as the level density in the barrier is much higher
than the one at the ground state deformation and therefore the considerations
of \cite{Otsuka25} might not apply here. One should also investigate
how the inertias change along the triaxial path when the tensor term is
included. Given the little impact on the axially symmetric inertias, one
could expect a rather limited impact and therefore the least action argument
will also apply here.  This is an interesting issue  but it seems to be 
out of the scope of the present study. We plan to analyze this issue in a forthcoming study.

The comparison between the excitation energies $E_{II,GCM}$ of the 
fission isomers and the  heights $B_{II,GCM}$ of the outer barriers 
also reveals a  Gogny-D1ST reduction in those quantities. For the 
first, we have obtained $E_{II,GCM}^{D1S}=1.27$ and 
$E_{II,GCM}^{D1ST}=0.63$~MeV, while for the second  the values are 
$B_{II,GCM}^{D1S}=2.44$ and $B_{II,GCM}^{D1ST}=1.84$~MeV. 

\begin{figure}
\includegraphics[width=0.42\textwidth]{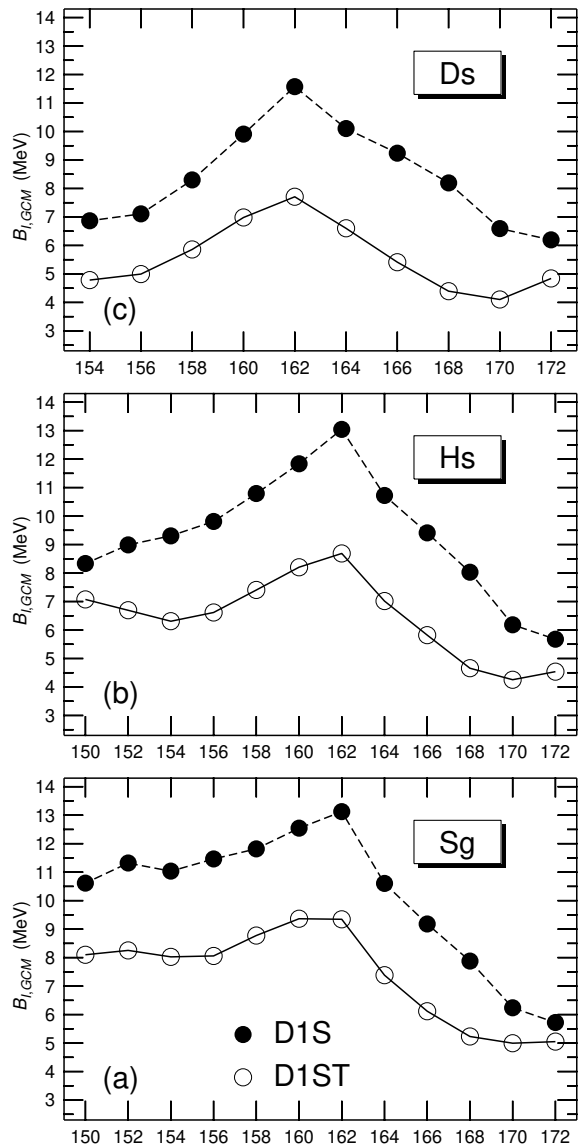}
\caption{Inner barrier heights $B_{I,GCM}$ for the nuclei 
$^{256-278}$Sg [panel (a)], $^{258-280}$Hs [panel (b)] and 
$^{264-282}$Ds [panel (c)], as functions of the neutron number. Note, 
that the $B_{I,GCM}$ values contain the effects associated with 
zero-point rotational and vibrational corrections. Results have been 
obtained with the parametrizations D1S and D1ST of the Gogny-EDF. For 
more details, see the main text.
}
\label{Inner_SgHsDs} 
\end{figure}

The reduction of the inner barrier height observed when going from D1S 
to D1ST is certainly desirable, as it might impact the predicted 
spontaneous fission half-lives in the right direction 
\cite{Fission-tensor-raynerJPG}. We have to point out that a similar 
reduction has been observed in a bunch of actinide nuclei in 
calculations with Gogny DG \cite{finite-range-tensor-refit-gogny}, a 
new form of the Gogny force with a tensor term and a finite range 
density dependent interaction. We stress, however, that a direct 
correlation between barrier heights and $t_{SF}$ values should be 
treated with caution \cite{MA-1,MA-3,MA-4,Inner-1}. First, for a given 
nucleus, the probability to penetrate the fission barrier depends not 
only on the height but also on the shape (mainly the width) of the 
barrier. Second, as can be seen from Eqs.(\ref{TSF}), (\ref{Action}) 
and  (\ref{integrand-eq}), the predicted $t_{SF}$ values are also 
affected by the behavior and size of the collective inertia 
$M_{GCM}(Q_{20})$. Therefore, besides the topography of the collective 
potential $V_{GCM}(Q_{20})$ in between the classical turning points, 
the specific modulation effects arising from $M_{GCM}(Q_{20})$ should 
be taken into account to make quantitative predictions. In turn, the 
behavior and size of the collective mass $M_{GCM}(Q_{20})$ is strongly 
influenced by pairing correlations 
\cite{Rayner-U-PRC,turning,Inv-Flocard}.

In panels (c) and (d) of Fig.~\ref{Pegagogival_260Rf}, we have depicted 
the proton and neutron pairing interaction energies $E_{pp,\tau} = 
\frac{1}{2} \textrm{Tr}(\Delta_{\tau} \kappa_{\tau} )$ (with $\tau=Z$ 
and $N$) \cite{Ringbook} for $^{260}$Rf. On the average, the D1ST 
proton and neutron pairing energies are slightly larger than the D1S 
ones up to the fission isomer. From there on, the D1ST ones are 
significantly larger. Here, one should keep in mind, that the tensor 
contribution to the pairing field is not taken into account
in the Gogny-D1ST EDF 
\cite{Marta-tensor-1,Marta-tensor-2,Marta-tensor-3,Marta-tensor-4,Marta-tensor-5, 
Fission-tensor-2020,Fission-tensor-raynerJPG}. Therefore, the proton 
and neutron pairing differences observed in the panels reflect the 
rearrangement of the proton and neutron single-particle energies due to 
the tensor term \cite{Ringbook}.

As can be seen from panel (c), the proton pairing interaction energies 
$E_{pp,Z}^{D1S}$ and $E_{pp,Z}^{D1ST}$ display minima at $Q_{20}=16$~b 
and around $Q_{20}=60$~b, in coincidence with the position of the 
ground state and fission isomer. In the case of $E_{pp,Z}^{D1ST}$, the 
minimum around $Q_{20}=60$~b is less pronounced and it is accompanied 
by two other local minima around $Q_{20}=50$~b and $Q_{20}=70$~b. Both 
$E_{pp,Z}^{D1S}$ and $E_{pp,Z}^{D1ST}$ exhibit maxima at $Q_{20}=36$~b  
and around $Q_{20}=72$~b, in coincidence with the position of the first 
and second fission barriers.

The neutron pairing interaction energies $E_{pp,N}^{D1S}$ and 
$E_{pp,N}^{D1ST}$ in panel (d) show minima at $Q_{20}=16$~b as well as 
around $Q_{20}=32$~b and $Q_{20}=50$~b. On the other hand, maxima are 
observed around $Q_{20}=26$~b and $Q_{20}=36$~b. The energy 
$E_{pp,N}^{D1ST}$ shows a peak around $Q_{20}=72$~b, while for 
$E_{pp,N}^{D1S}$ such a peak is shifted to a larger quadrupole 
deformation.

In panels (a) and (b) [(c) and (d)] of Fig.~\ref{spes_260Rf}, we have 
plotted the D1S (D1ST) proton and neutron single-particle energies  for 
$^{260}$Rf, as functions of the quadrupole deformation $Q_{20}$. In 
order to obtain those Nilsson-like single-particle diagrams, we have 
computed the eigenvalues of the Routhian 
$h=t+\Gamma-\lambda_{Q_{20}}Q_{20}$, with $t$ being the kinetic energy, 
$\Gamma$ the Hartree-Fock field and $\lambda_{Q_{20}}Q_{20}$ the term 
containing the Lagrange multiplier to enforce the constrain in the 
quadrupole moment. As it is well known, regions with low densities of 
single-particle orbitals (Jahn-Teller effect) in those diagrams tend to 
favor the existence of minima as functions of the quadrupole 
deformation $Q_{20}$. Vertical dashed lines have been plotted to signal
the position of the ground state (GS), inner barrier (IB) and fission
isomer (FI). Finally, the thick dotted line represents the Fermi level.

We observe in the plot that nearly the same relevant orbitals are 
present in the energy window in both calculations, but their relative 
position is different at the spherical configuration. The evolution 
with quadrupole deformation is not the same in the two cases, but 
energy gaps emerge at the position of the ground state and fission 
isomer in both the D1S and D1ST calculations. For example, the D1S and 
D1ST proton single-particle energies shown in panels (a) and (c) 
display  energy gaps around $Q_{20}=16$~b and $Q_{20}=56$~b that favor 
a deformed ground state and the fission isomer due to the Jahn-Teller 
effect. To quantify this effect one case use the energy level density
defined in \cite{Bernard23} for protons and neutrons $\eta^{p,n}$. 
For  $Q_{20}=16$~b one gets $\eta^{p}=1.34$ for D1S and a similar value 
of $\eta^{p}=1.28$ for D1ST.  In the  neutron single-particle spectra in panels (b) and 
(d) around $Q_{20}=16$~b one observes a large level density around the Fermi level that is not
favoring the formation of the ground state minimum. 
In this case, one obtains $\eta^{n}=3.48$ for D1S and a similar value 
of $\eta^{n}=3.66$ for D1ST, values which are significantly larger than the proton
case. Furthermore, both the D1S and D1ST proton and neutron 
single-particle spectra, exhibit a mild gap around $Q_{20}=30$~b. Such a 
gap, which is more pronounced in calculations with the D1S EDF, 
provides a mechanism for the emergence of a shoulder between the ground 
state and the top of the inner barrier [see, panel (a) of 
Fig.~\ref{Pegagogival_260Rf}].  As will be discussed later in this 
paper (see, Sec.~\ref{systematic-FP}), with increasing neutron number 
this shoulder gradually evolves, leading to a progressive fragmentation 
of the inner barrier and eventually developing into a local prolate 
minimum. There is also a gap at $Q_{20}=60$~b in the neutron single-particle 
spectrum responsible for the fission isomer minimum. 
 
As expected, in the region around the top of the fission barrier, the 
single-particle spectrum shows a high level density  both in the proton 
and neutron case. However, such high level density makes difficult the 
identification of the mechanism explaining why the fission barrier 
heights obtained with D1ST are smaller than the ones obtained with D1S. 
To give an idea of the complexity, let us mention that in the neutron 
energy window of Fig.~\ref{spes_260Rf} there are 25 neutron levels at 
$Q_{20}=36$~b (top of the fission barrier) in both D1S and D1ST cases. 
The orbitals are  essentially the same (same $K$, parity, dominant 
Nilsson quantum number) but they are located at slightly different 
energies. For this quadrupole moment one gets $\eta^{n}=2.76$ for D1S 
and a similar value of $\eta^{n}=2.85$ for D1ST. On the other hand, 
$\eta^{p}=4.19$ for D1S and a  value of $\eta^{p}=3.54$ for D1ST. From 
the values of the level densities one can conclude that for the inner barrier, the 
role of protons and neutrons is the same as in the ground state: 
protons favor the maximum of the inner barrier. It also not so evident 
that the different values $\eta^{p}=4.19$ for D1S and $\eta^{p}=3.54$ 
for D1ST are enough as to explain the lower inner barrier height of D1ST. We 
conclude that the inclusion of the tensor term lowers the fission 
barrier height through a collective effect, in which contributions from 
many orbitals are involved.

Coming back to Fig.~\ref{Pegagogival_260Rf}, in panel (e) we have 
plotted the D1ST and D1S GCM collective inertias $M_{GCM}(Q_{20})$ as  
functions of the quadrupole moment. Let us mention, that in all the 
computations of the $t_{SF}$ values, the wiggles in the collective 
masses have been softened by means of a three point filter 
\cite{Rayner-U-PRC}. Both GCM masses display a similar qualitative 
trend, which is well correlated with the patterns found in the proton 
and neutron pairing and single-particle energies. We have found that 
for $10~\textrm{b} \le Q_{20} \le 20~\textrm{b}$ the D1ST mass is 
slightly larger than the D1S one, while for larger quadrupole moments 
the former is, on the average, smaller than the latter. The differences
found in the two masses are not significant enough as to warrant a 
substantial change in the $t_{SF}$ values.

We have computed the 
corresponding D1S and D1ST spontaneous fission half-lives 
for $^{260}$Rf (with $E_{0}=0.5$ MeV \cite{Fission-tensor-raynerJPG})
and obtained the values
$ \log_{10}~t_{SF,GCM}^{D1S}=5.51$  and  $\log_{10}~t_{SF,GCM}^{D1ST}=-1.08$ with $t_{SF}$ in seconds.
They should be compared with the experimental value  $\log_{10}~t_{SF}^{EXP}=-1.69$ \cite{exp-tsf}. 
The striking quenching of $t_{SF}$ in the calculation with the tensor term
is due to the reduction of the first barrier height and is the main motivation
to extend the calculations to other even-even superheavy nuclei with $Z > 100$.
It is very satisfying to notice also the improved agreement with the experiment.

It is important to note that for $^{260}$Rf, $Q_{20}=88$~b represents 
the last quadrupole configuration contributing to the spontaneous 
fission half-lives obtained with both the D1S and D1ST EDFs. Therefore, 
the outer barriers still contribute to the corresponding $t_{SF}$ 
values [see also, panel (a) of Fig.~\ref{FissionPAths_Rf}]. If these 
outer barriers are neglected -for instance, by adopting $Q_{20}=60$~b 
as the last contributing configuration- the resulting values are $ 
\log_{10}~t_{SF,GCM}^{D1S}=0.21$  and  
$\log_{10}~t_{SF,GCM}^{D1ST}=-4.03$. Such {\it{ad hoc}} choices, 
however, are inconsistent with the determination of the classical 
turning points based on the microscopic collective potential for a 
given value of the parameter $E_{0}$, and have not been adopted in the 
present study.

Most of the results explicitly  discussed in this paper correspond to 
collective potentials $V(Q_{20})$ and masses $M(Q_{20})$ computed 
within the GCM scheme. However, we have also determined such quantities 
using the ATDHFB approximation. As representative results of our 
calculations, panels (a) and (b) of Fig.~\ref{comp-GCM-ATD} compare the 
GCM and ATDHFB collective potentials and masses obtained for $^{260}$Rf 
using the D1S and D1ST EDFs.

On the one hand, the potentials and masses in Fig.~\ref{comp-GCM-ATD}, 
exhibit a rather similar  behavior as functions of the quadrupole 
deformation, with maxima and minima located nearly at the same position 
in both cases. On the other hand, regardless of the EDF employed in the 
calculations, the ATDHFB inner barrier heights, excitation energies of 
fission isomers and outer barrier heights are smaller than the GCM 
values. Nevertheless, as can be seen from panel (b), the ATDHFB masses 
are much larger than the GCM ones 
\cite{SH-d1mstar,Rayner-U-PRC,Fission-tensor-raynerJPG}. The increase 
in the mass $M_{ATDHFB}(Q_{20})$ compensates the ATDHFB reductions in 
barrier heights and excitation energies of fission isomers, as compared 
to the corresponding GCM values. As a consequence, the ATDHFB values 
for   spontaneous fission half-lives are larger than the corresponding 
GCM values. For example, in the case of $^{260}$Rf, we have obtained $ 
\log_{10}~t_{SF,ATDHFB}^{D1S}=7.08$ and 
$\log_{10}~t_{SF,ATDHFB}^{D1ST}=0.37$, which should be compared with the 
values $ \log_{10}~t_{SF,GCM}^{D1S}=5.51$, 
$\log_{10}~t_{SF,GCM}^{D1ST}=-1.08$ and $\log_{10}~t_{SF}^{EXP}=-1.69$ 
\cite{exp-tsf}. Similar results have been obtained for all the studied 
nuclei.

\subsection{Systematic of the fission paths and spontaneous 
fission half-lives  in No, Rf, Sg, Hs and Ds nuclei}
\label{systematic-FP} 

Let us turn our attention to the discussion of the main features 
observed along the fission paths of  $^{254-270}$Rf, $^{256-278}$Sg and 
$^{258-280}$Hs. Similar results have been obtained for $^{250-266}$No 
and $^{264-282}$Ds.

The D1S and D1ST GCM collective potentials, obtained for the isotopes 
$^{254-270}$Rf, are plotted in panels (a) and (b) of 
Fig.~\ref{FissionPAths_Rf}. For all the considered isotopic chains, the 
excitation energies $\delta E_{sph,GCM}$ of the spherical 
configurations decrease with increasing neutron number. For example, we 
have obtained $\delta E_{sph,GCM}^{D1S}=$ 23.95, 22.02, 19.56, 12.98~MeV 
and $\delta E_{sph,GCM}^{D1ST}=$ 19.51, 18.80, 17.03, 
11.48~MeV for $^{258,262,266,270}$Rf.

Regardless of the employed EDF, the absolute minima of the paths in 
Fig.~\ref{FissionPAths_Rf} are located at $Q_{20}=14-16$~b. For 
$^{254}$Rf, the top of the inner  barrier is located at $Q_{20}=32$~b. 
However, as the neutron number increases, the location of the top of the barrier 
shifts, attaining  $Q_{20}=40$~b in $^{270}$Rf. A shoulder emerges 
between the ground state and the top of the inner barrier, becoming a 
shallow prolate local minimum at $Q_{20}=28$~b in  $^{268,270}$Rf. Due 
to the gradual fragmentation of the inner barrier, in these and similar 
cases we refer to the largest inner barrier height in our discussion 
\cite{SH-d1mstar}.
 
The reduction of the D1ST inner barrier heights  $B_{I,GCM}^{D1ST}$, as 
compared with the D1S $B_{I,GCM}^{D1S}$ values, in Rf isotopes is 
clearly seen from Fig.~\ref{FissionPAths_Rf}. Those barrier heights are 
depicted in panel (a) of Fig.~\ref{Inner_Rf_No}, as functions of the 
neutron number $N$. The differences $\delta B_{I,GCM} = 
B_{I,GCM}^{D1S}- B_{I,GCM}^{D1ST}$ are within the range 
$2.5~\textrm{MeV} \le \delta B_{I,GCM} \le 3.4~\textrm{MeV}$. 
Regardless of the EDF employed in the calculations, the barrier heights 
show peaks at $N=152$ and $N=160-162$. Note, however, that the D1ST 
peaks are less pronounced than the D1S ones. On the other hand, the 
barrier heights decrease substantially for both $^{268,270}$Rf.

The inner barrier heights obtained for $^{250-266}$~No are shown in 
panel (b) of Fig.~\ref{Inner_Rf_No}. We have obtained differences 
within the range $2.2~\textrm{MeV} \le \delta B_{I,GCM} \le 
3.7~\textrm{MeV}$.  In this case, the barrier heights reach their 
largest values for $N=150$, two units less than in the Rf case. The origin
of this difference can be attributed to the fact that the barriers include
zero point energy corrections which strongly depend on pairing properties and
are not so correlated with shell effects.  It is worth noting that, within the neutron 
number range $150 \le N \le 162$, the D1S barrier heights display an 
approximately linear decrease as a function of $N$. In contrast, the 
D1ST barrier heights show a plateau-like behavior. With both EDFs the 
barrier heights decrease substantially for $^{266}$No. 

\begin{figure*}
\includegraphics[width=1.\textwidth]{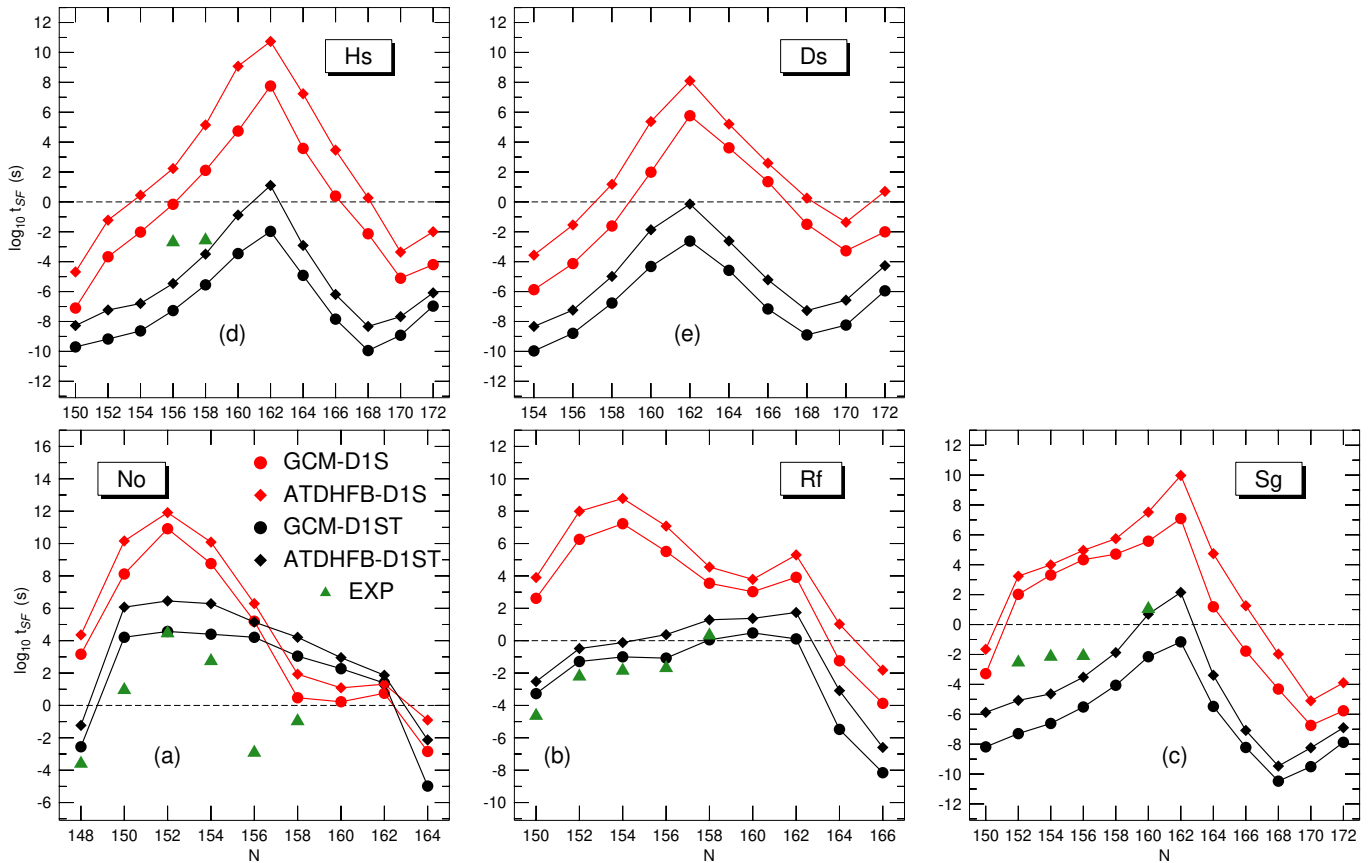}
\caption{(Color online) The spontaneous fission half-lives t$_{SF}$, predicted for the nuclei
$^{250-266}$No [panel (a)], $^{254-270}$Rf [panel (b)], $^{256-278}$Sg [panel (c)], 
$^{258-280}$Hs [panel (d)] and $^{264-282}$Ds [panel (e)], within the GCM
and ATDHFB schemes, are depicted as functions of the neutron number. Results have been obtained 
with the parametrizations D1S and D1ST of the Gogny-EDF and $E_{0}$ = 0.5 $MeV$. The available 
experimental values, taken from Ref.~\cite{exp-tsf}, are included in the plots. 
For more details, see the main text.
}
\label{tsf-summary} 
\end{figure*}

Coming back to Fig.~\ref{FissionPAths_Rf}, we  observe shallow fission 
isomers for the lighter Rf isotopes. These (reflection symmetric) 
fission isomers disappear with increasing neutron number $N$. Their 
excitation energies are reduced in the Gogny-D1ST calculations. For 
example, we have obtained the values $E_{II,GCM}^{D1S}=1.11$, 1.96, 
1.51, 1.27 MeV and $E_{II,GCM}^{D1ST}=0.06$, 0.48, 0.54, 0.63~MeV 
for $^{254-260}$Rf. The outer barriers play a role in the spontaneous 
fission properties of the lighter Rf isotopes but also disappear with 
increasing $N$. For example, we have obtained the heights 
$B_{II,GCM}^{D1S}=3.13$, 4.20, 3.84, 2.44~MeV and 
$B_{II,GCM}^{D1ST}=1.70$, 2.03, 1.96, 1.84~MeV for $^{254-260}$Rf.

As can be seen from Fig.~\ref{FissionPAths_Rf}, the outer sectors of 
the D1ST collective potentials in Rf isotopes with $N \ge 158$ display 
a gentler decline than the D1S ones. However, for those Rf isotopes,  
this does not have an impact on the predicted $t_{SF}$ values.
 
A similar gentler decline is found in the outer sectors of the D1ST 
paths for No isotopes with $N \ge 158$. Nevertheless, in this case, 
this feature plays a more subtle role for nuclei such as $^{260-264}$No 
[see, panel (a) of  Fig.~\ref{tsf-summary}]. For example, in the case 
of $^{260}$No, the Gogny-D1ST calculations predict an outer barrier, 
with the top at $Q_{20}=70$~b and the height $B_{II,GCM}^{D1ST}=1.98$ 
MeV. For this nucleus,  $Q_{20}=90$~b represents the last quadrupole 
configuration contributing to the spontaneous fission half-life 
$\log_{10}~t_{SF,GCM}^{D1ST}=3.04$, i.e., the outer barrier still plays 
a role. On the other hand, due to the faster decline in the outer 
sector of the D1S potential, the outer barrier disappears and 
$Q_{20}=68$~b is the last configuration contributing to the smaller 
value $\log_{10}~t_{SF,GCM}^{D1S}=0.48$. Therefore, for $^{260}$No the 
gentler decline in the outer sector of the D1ST potential, as compared 
to the D1S one, still compensates the reduction of 2.47~MeV  in the 
inner barrier height and leads to a larger $t_{SF}$ value. The same 
arguments can be applied  for $^{262,264}$No.

The D1S and D1ST GCM collective potentials obtained for $^{256-278}$Sg 
and $^{258-280}$Hs are shown in panels (a) and (b) of 
Figs.~\ref{FissionPAths_Sg} and \ref{FissionPAths_Hs}. The ground 
states in Sg and Hs isotopes up to around  $N=166$ correspond to 
$Q_{20}=14-18$~b. On the other hand, in the more neutron-rich sectors 
of the chains, as one moves toward the neutron  shell closure $N=184$ 
\cite{SH-d1mstar}, the ground state deformations are reduced up to 
$Q_{20}=10$~b.

Broad inner fission barriers are obtained in the D1S and D1ST 
calculations for  Sg and Hs nuclei. For $^{256}$Sg and $^{258}$Hs, the 
top of the inner  barrier is located at $Q_{20}=32$~b. However, as the 
neutron number increases, the location of the top of the barrier 
shifts, attaining  $Q_{20}=44-46$~b in $^{276,278}$Sg and $^{280}$Hs. 
The fragmentation of the inner barrier is also apparent in the Sg and 
Hs isotopic chains as a result of the emergence of a shoulder, which 
becomes a local prolate minimum around $Q_{20}=28-30$~b in heavier  
isotopes. This is also the case for Ds nuclei.
 
The largest inner barrier heights $B_{I,GCM}^{D1S}$ and 
$B_{I,GCM}^{D1ST}$ obtained for $^{256-278}$Sg,$^{258-280}$Hs and 
$^{264-282}$Ds are shown in panels (a), (b) and (c) of 
Fig.~\ref{Inner_SgHsDs}, as functions of the neutron number $N$. 
Regardless of the EDF employed in the calculations, the  heights in all 
three isotopic chains exhibit a similar dependence on neutron number, 
characterized by pronounced maxima at $N=162$. However, from 
Figs.~\ref{FissionPAths_Sg}, \ref{FissionPAths_Hs} and 
\ref{Inner_SgHsDs} one realizes that the inclusion of tensor 
correlations also leads to a  significant reduction of the inner 
barrier heights in these isotopic chains. We have obtained differences 
$\delta B_{I,GCM}$ within the ranges $0.67~\textrm{MeV} \le \delta 
B_{I,GCM} \le 3.77~\textrm{MeV}$, $1.14~\textrm{MeV} \le \delta 
B_{I,GCM} \le 4.35~\textrm{MeV}$ and  $1.35~\textrm{MeV} \le \delta 
B_{I,GCM} \le  3.87~\textrm{MeV}$  for Sg, Hs and Ds nuclei, 
respectively, with the smallest values corresponding to the $N=172$ 
isotones $^{278}$Sg,  $^{280}$Hs and  $^{282}$Ds.

For Sg isotopes, $Q_{20}$-configurations associated with fission 
isomers and/or  outer barriers play no role in the determination of the 
D1ST $t_{SF}$ values. This is also the case for Hs and Ds nuclei 
regardless of the EDF employed in the calculations. Note, that with 
increasing atomic number $Z$ the collective potentials exhibit an 
overall faster decline. On the other hand, in calculations with the 
Gogny-D1S EDF for lighter Sg isotopes both fission isomers and outer 
barriers still play a role in the determination of the actual 
lifetimes. As a typical example, let us consider the D1S GCM collective 
potential for the nucleus $^{260}$Sg [see, panel (a) of  
Fig.~\ref{FissionPAths_Sg}]. Such a potential exhibits a fission isomer 
at $Q_{20}=64$~b with an excitation energy $E_{II,GCM}^{D1S}=0.50$~MeV, 
while the top of the outer barrier is located at $Q_{20}=74$~b  with 
the height $B_{II,GCM}^{D1S}=2.18$ MeV. For this nucleus,  
$Q_{20}=82$~b represents the last quadrupole configuration contributing 
to the spontaneous fission half-life $ \log_{10}~t_{SF,GCM}^{D1S}=3.31$. 
As can be seen from Figs.~\ref{FissionPAths_Sg} and 
\ref{FissionPAths_Hs}, the outer sectors of the D1ST collective 
potentials in heavier isotopes display a gentler decline than the D1S 
ones. However, such a feature does not affect the predicted $t_{SF}$ 
values for Sg and Hs nuclei. Similar results have been obtained for Ds 
isotopes.

The GCM spontaneous fission half-lives, obtained for the No, Rf, Sg, Hs 
and Ds isotopic chains are depicted in panels (a)-(e) of 
Fig.~\ref{tsf-summary}, as functions of the neutron number. We have 
adopted the typical value  $E_{0} = 0.5$ MeV in the calculation 
\cite{Fission-tensor-raynerJPG}. For the No isotopes, panel (a) shows 
that the Gogny-D1S calculations predict $t_{SF,GCM}^{D1S}$ values 
exhibiting a pronounced maximum at $N=152$ and that, for 
$^{250-256}$No, substantially overestimate the experimental data 
\cite{exp-tsf}. In contrast, the inclusion of tensor correlations 
within the Gogny-D1ST HFB framework substantially reduces the  
lifetimes for $^{250-256}$No, leading to $t_{SF,GCM}^{D1ST}$ values in 
better agreement with the experiment. It is also worth emphasizing that 
tensor correlations modify the overall profile of the D1ST half-lives 
relative to the D1S results. Although the D1ST half-lives still attain 
their largest value at $N=152$, they display a plateau-like behavior 
for $^{252-256}$No. To some extent, these differences in the 
$t_{SF,GCM}^{D1S}$ and $t_{SF,GCM}^{D1ST}$ profiles, reflect the 
differences observed in panel (b) of Fig.~\ref{Inner_Rf_No} for the 
profiles of the inner barrier heights. Regardless of the employed EDF, 
the sudden drop of $\log_{10}~t_{SF}^{EXP}$ for $^{258}$No 
\cite{exp-tsf} is not reproduced by the calculations, a deficiency
already noted in previous studies \cite{SH-d1mstar,MA-4,MF-fission-6}, 
based on different parametrizations of the Gogny-EDF (without tensor 
terms). On the other hand, for $^{260-264}$No the predicted 
$t_{SF,GCM}^{D1ST}$ values are larger than the $t_{SF,GCM}^{D1S}$ ones. 
As already discussed, for those nuclei this is a consequence of the 
gentler decline in the outer sectors of the D1ST potentials, as 
compared to the D1S ones, that still compensates the reduction in the 
D1ST inner barrier heights. The small kink for $^{264}$No observed in 
the Gogny-D1S calculation disappears in the Gogny-D1ST case. 
Nevertheless, both EDFs predict a reduction of the half-life for 
$^{266}$No. 

As can be seen from panel (b), the largest $t_{SF,GCM}^{D1S}$ value 
corresponds to $N=154$. A kink, more pronounced than the one observed 
in the No isotopic chain, is obtained for $N=162$ with the Gogny-D1S 
EDF. The predicted $t_{SF,GCM}^{D1S}$ lifetimes significantly 
overestimate the experimental data \cite{exp-tsf} and do not reproduce 
the increase in $\log_{10}~t_{SF}^{EXP}$  in going from $^{254}$Rf to 
$^{262}$Rf \cite{SH-d1mstar}. However, a drastic change is found in the 
half-life profile within the Gogny-D1ST HFB framework. In particular, 
the peak at $N=154$ in the calculation with D1S disappears in the D1ST 
case, while the kink at $N=162$ is softened and it becomes the global 
maximum of the $t_{SF,GCM}^{D1ST}$ lifetimes. To some extent, the 
differences in the $t_{SF,GCM}^{D1S}$ and $t_{SF,GCM}^{D1ST}$ profiles, 
reflect the differences observed in panel (a) of Fig.~\ref{Inner_Rf_No} 
for the profiles of the inner barrier heights. The change in the D1ST 
profile, as compared with the D1S one, is accompanied by a significant 
reduction in the predicted  $t_{SF,GCM}^{D1ST}$ values leading to an 
almost quantitative agreement with the available experimental data for 
$^{254-262}$Rf. Both Gogny-EDFs predict a pronounced decrease in the 
spontaneous fission half-lives beyond $N=162$.

The $t_{SF,GCM}^{D1S}$ and $t_{SF,GCM}^{D1ST}$ half-lives, shown in 
panels (c)-(e) for the Sg, Hs, and Ds isotopes, display a similar 
dependence on neutron number, with pronounced maxima at $N = 162$. This 
trend closely mirrors that of the inner barrier heights shown in panels 
(a)-(c) of Fig.~\ref{Inner_SgHsDs}. It arises from the more rapid 
decrease of the collective potentials with increasing atomic number 
$Z$, which -apart from the lighter Sg isotopes in the Gogny-D1S 
calculations- results in spontaneous fission properties solely governed 
by the properties of the inner  barriers. Once more, one sees that 
tensor correlations provide $t_{SF,GCM}^{D1ST}$ half-lives much smaller 
than their D1S counterparts. It is worth noting that the  predicted GCM 
$t_{SF,GCM}^{D1S}$ lifetimes overestimate considerably the available 
$\log_{10}~t_{SF}^{EXP}$ data \cite{exp-tsf} for Sg and Hs nuclei, while 
the corresponding $t_{SF,GCM}^{D1ST}$ values underestimate those data.

For all the considered nuclei, we have also evaluated the spontaneous 
fission half-lives within the ATDHFB scheme 
\cite{Rayner-U-PRC,collective-massesSamuel} and $E_{0} = 0.5$ MeV 
\cite{Fission-tensor-raynerJPG}. The corresponding D1S and D1ST 
$t_{SF,ATDHFB}$ values are also included in panels (a)-(e) of 
Fig.~\ref{tsf-summary}. It is satisfying to note that, for both the D1S 
and D1ST EDFs, the GCM and ATDHFB predictions follow the same trend, 
although the ATDHFB lifetimes are systematically larger than the GCM 
ones. The larger ATDHFB half-lives stem from the  larger collective 
masses obtained  within the ATDHFB framework 
\cite{Rayner-U-PRC,collective-massesSamuel,Fission-tensor-raynerJPG} 
(see also, Sec.~\ref{methodology-260Rf}). These results confirm the 
robustness of the $t_{SF}$ trends obtained in this work  for No, Rf, 
Sg, Hs and Ds nuclei with respect to the scheme employed to compute the 
collective potentials and masses. Incidentally, it is interesting to 
note that at least for some of the studied Sg and Hs isotopes, the 
predicted ATDHFB half-lives agree better with the experimental ones 
\cite{exp-tsf}.

\section{Conclusions} 
\label{conclusions} 
In this work we have considered the impact of introducing 
 a perturbative tensor term in  
the Gogny D1S interaction (the D1ST parametrization) when it is used 
to describe fission properties of a set of even-even superheavy nuclei from No (Z=102) 
to Ds (Z=110). It is observed that pairing properties of D1S do not change 
significantly when the tensor term is introduced and, as a consequence, 
the impact on the collective inertia for the quadrupole collective 
degree of freedom is minor. On the other hand, it is noticed that the 
quadrupole moment of the most relevant configurations in the collective 
potential (ground state, inner barrier, fission isomer and second 
barrier) is essentially independent of the interaction used. In 
addition, the fission barrier heights obtained with D1ST are smaller 
than those obtained with D1S while the size of the collective inertias remains more or
less the same. As a consequence, the spontaneous fission half-lives 
obtained with D1ST are significantly shorter than those obtained with 
D1S. The reduction helps improving the poor agreement of D1S' fission 
lifetimes as compared with the experimental data in the superheavy 
region (Z > 100) of the nuclear chart. Therefore, we conclude that the tensor 
interaction represents  an important ingredient in the microscopic 
description of fission and its impact cannot be overlooked. We have restricted
ourselves to nuclei below Z=110 for practical reasons. Given the impact
of the tensor term in the fission barriers we plan to extend the present
study to nuclei in the region Z=112-120 including very neutron rich isotopes. We also 
recognize the importance of using an interaction with a tensor term 
that is fully integrated in the fitting protocol as the recent proposal 
of Ref.~\cite{finite-range-tensor-refit-gogny}. Work in this direction 
is in progress and will be reported in the near future.

\begin{acknowledgments} 
The work of R. Rodr\'{\i}guez-Guzm\'an and  A. Rakhmankulov 
is funded by Nazarbayev University under the Faculty
Development Competitive Research Grants Program (FDCRGP) for
2025-2027, Grant 040225FD4712 and under the Collaborative
Research Program (CRP) for 2026-2028, Grant 110326CRP0803. The 
work of LMR is supported by Spanish Agencia Estatal de Investigacion 
(AEI) of the Ministry of Science and Innovation under Grant No. 
PID2021-127890NB-I00 and PID2024-159559NB-C21. Both R.R.-G. and A.R. 
also acknowledge the support provided by the IT Department for 
performing calculations on the HPC clusters Shabyt, Irgetas and Muon at 
Nazarbayev University.
\end{acknowledgments}

\section{Data availability}

The data that support the findings of this study are available upon reasonable request from the authors.

%
%

\end{document}